\documentclass[aps,prx,twocolumn,superscriptaddress,longbibliography]{revtex4-1}
\usepackage[utf8]{inputenc}

\usepackage{graphicx}% Include figure files
\usepackage{epstopdf}
\usepackage{bm}% bold math
\usepackage{amsmath}
\usepackage{verbatim}
\usepackage{amssymb}
\usepackage{listings}
\usepackage{color}
\definecolor{darkblue}{rgb}{0,0,0.6}
\usepackage{here}
\usepackage[colorlinks,linkcolor=darkblue,citecolor=darkblue,urlcolor=darkblue]{hyperref}
\renewcommand\vec[1]{\boldsymbol{#1}}

\newcommand\Fig[1]{Fig.~\ref{#1}}

\newcommand\App[1]{Appendix~\ref{#1}}

\renewcommand\vec[1]{\boldsymbol{#1}}
\newcommand\eq[1]{Eq.~(\ref{#1})}
\newcommand{\rev}[1]{\textcolor{black}{#1}}
\newcommand{\revv}[1]{\textcolor{black}{#1}}

\newcommand\bigang[1]{\left\langle #1 \right\rangle}
\newcommand\bigpar[1]{\left( #1 \right)}

\begin{document}

\title{Relaxation dynamics in the energy landscape of glass-forming liquids}

\author{Yoshihiko Nishikawa}

\affiliation{Laboratoire Charles Coulomb (L2C), Universit\'e de Montpellier, CNRS, 34095 Montpellier, France}

\author{Misaki Ozawa}

\affiliation{Laboratoire de Physique de l'Ecole normale sup\'erieure, ENS, Universit\'e PSL, CNRS, Sorbonne Universit\'e, Universit\'e Paris-Diderot, Sorbonne Paris Cit\'e, Paris, France}

\author{Atsushi Ikeda}

\affiliation{Graduate School of Arts and Sciences, The University of Tokyo, Tokyo 153-8902, Japan}

\author{Pinaki Chaudhuri}

\affiliation{The Institute of Mathematical Sciences, C.I.T. Campus, Taramani, Chennai 600 113, India}

\author{Ludovic Berthier}

\affiliation{Laboratoire Charles Coulomb (L2C), Universit\'e de Montpellier, CNRS, 34095 Montpellier, France}

\affiliation{Yusuf Hamied Department of Chemistry, University of Cambridge, Lensfield Road, Cambridge CB2 1EW, United Kingdom}

\date{\today}

\begin{abstract}
  We numerically study the \rev{zero-temperature} relaxation dynamics of several glass-forming models to their inherent structures, following quenches from equilibrium configurations sampled across a wide range of \rev{initial} temperatures. In a mean-field Mari-Kurchan model, we find that relaxation changes from a power-law to an exponential decay below a well-defined temperature, consistent with recent findings in mean-field $p$-spin models. By contrast, for finite-dimensional systems, the relaxation is always algebraic, with a non-trivial universal exponent at high temperatures crossing over to a harmonic value at low temperatures. We demonstrate that this apparent evolution is controlled by a temperature-dependent population of localised \rev{glassy excitations. Our work unifies several recent lines of studies aiming at a detailed characterisation of the complex potential energy landscape of glass-formers, and challenges both mean-field and real space descriptions of glasses.}
\end{abstract}

\maketitle

\section{Introduction}

Many systems of scientific interest are described as `complex', even though definitions of complexity may vary across scientific fields~\cite{bouchaud2011complex}. For many-body interacting systems, the potential energy landscape, $E(\{ {\vec r} \})$, which describes the potential energy $E$ of the system as a function of the complete set of coordinates $\{ {\vec r} \}$ of its constituents, has become a central object of study~\cite{wales2004energy,stillinger2015energy}. It serves both empirical goals, for instance to picture the dynamic evolution of a system in a `rugged' landscape~\cite{stillinger1995topographic}, but can also be described mathematically very precisely~\cite{wales2004energy,auffinger2013random,ros2019complex}. The detailed characterisation and dynamic exploration of complex potential energy landscapes are important problems for \rev{amorphous materials~\cite{stillinger1995topographic,sciortino2005potential,heuer2008exploring}, optimisation problems~\cite{krzakala2007gibbs}, machine learning algorithms~\cite{lecun2015deep,baity2019comparing}, and other disordered systems~\cite{kent2021complex}.} 

Since the work of Goldstein~\cite{goldstein1969viscous}, the physics of glassy systems is often described in terms of the properties of their potential energy landscapes. The large number of energy minima connected by complex dynamic pathways is typically invoked in introductory lectures about amorphous media~\cite{stillinger1995topographic}, \rev{and the sketch of complex energy landscapes very often accompanies the interpretation of experimental measurements~\cite{angell1995formation}, which makes this object more than a pure theoretical curiosity.} Analytically, the properties of the potential energy landscape of glass-forming models have been studied extensively at the mean-field level through the analysis of fully-connected disordered spin models, such as $p$-spin models. In this limit, the phase space can be divided into long-lived metastable states \rev{(or, pure states),} and both free-energy and energy landscapes can be studied in great detail, thus providing a firm relation between the landscape structure and the thermodynamics and dynamics of the system~\cite{auffinger2013random,ros2019complex,cavagna1998stationary}. Current efforts in this area concern the analysis of dynamic pathways~\cite{ros2019complexity}, or corrections to mean-field~\cite{rizzo2020path}.

In finite dimensions, the study of energy minima, or inherent structures, first gained momentum when Stillinger and Weber transformed Goldstein's ideas into concrete tools to both explore and exploit the potential energy landscape of glasses~\cite{stillinger1995topographic,Stillinger1982}. A key step is the tiling of the equilibrium configuration space, pertinent to describe physical properties, into basins of attraction surrounding energy minima. It is this mapping which putatively connects the thermodynamic and dynamic properties of glass-formers to the topography of their potential energy landscape, although the relevance of such an approach has often been debated~\cite{berthier2003nontopographic,dyre2006colloquium}, because \rev{the pure states defined in the mean-field limit} do not exist in finite dimensions~\cite{Biroli2000}. The analysis of energy minima has been used to estimate the configurational entropy~\cite{sciortino2005potential}, while saddle points were discussed in connection with the dynamic mode-coupling crossover~\cite{cavagna2001fragile,angelani2000saddles,broderix2000energy,Grigera2002}. However, these approaches do not have the same level of rigour as those in $p$-spin models since inherent structures are different from pure states~\cite{Biroli2000,Berthier2014,Ozawa2018a}: \rev{inherent structures are configurations that are energetically stable against infinitesimal particle moves whereas the pure states are defined as free energy minima.} The structure of the potential energy landscape and its precise relationship with dynamics and thermodynamics remain under intense scrutiny~\cite{heuer2008exploring,baity2021revisiting}. In particular, the role of excitations in the potential energy landscape has been discussed in connection with sound propagation~\cite{gelin2016anomalous}, specific heat~\cite{khomenko2020depletion}, vibrational~\cite{lerner2016statistics} and mechanical properties~\cite{richard2020predicting}. 

Virtually all studies of glassy landscapes start by `instantaneously' relaxing configurations to the `nearest' energy minimum, following known numerical recipes~\cite{press1996numerical}. \revv{Strangely, however, only few studies have been dedicated to the physical processes at play during the energy minimization itself~\cite{Chacko2019,Folena2020,Gonzalez-Lopez2020,charbonneau2021memory,folena2021gradient,stanifer2021avalanche,manacorda2022gradient}.} \rev{In our view, this represents an important vacuum because this relaxation dynamics in fact provides a convenient way to navigate the potential energy landscape, explore its geometry and the nature as well as interactions between excitations that are relevant to describe glassy materials.} Suppose for instance that the landscape is simple and smooth. Using steepest descent dynamics, the system should then settle in an inherent structure very quickly while, on a rugged landscape, the system meanders and crosses many saddles during relaxation~\cite{Kurchan1996}. \rev{Similarly, the steepest descent dynamics obtained within kinetically constrained lattice models simply stems from a non-interacting set of excited defects and is therefore unremarkable~\cite{berthier2003nontopographic,berthier2003real}.} Thus, \rev{in the context of glassy systems}, the steepest descent dynamics probes the detailed structure of the potential energy landscape, potentially illuminates its connection to the physical dynamics, \rev{and provides novel constraints on physical descriptions of glassy excitations.}

Recently, the analysis of steepest descent in mixed mean-field $p$-spin glass models revealed the existence of two important characteristic temperatures~\cite{Folena2020}. First, starting from initial states prepared at high temperatures $T$, the energy density of the final inherent state is constant for $T > T_{\rm onset}$, and it decreases with decreasing $T$ when $T \leq T_{\rm onset}$. This sharp onset temperature does not affect the relaxation dynamics itself which obeys a non-trivial power-law time dependence as long as $T \geq T_{\rm SF}$. By contrast, the decay is exponentially fast below $T_{\rm SF}$ (initials stand for `State Following'). This implies that the system is always close to an energy minimum for $T \leq T_{\rm SF}$ in which it converges very quickly by steepest descent. The critical temperature $T_{\rm SF} < T_{\rm onset}$ also reflects a change in the structure of the potential energy landscape, as inherent states have a marginal density of states above $T_{\rm SF}$, which becomes gapped below. Within $p$-spin models, these two characteristic temperatures are unrelated to the equilibrium dynamics, which becomes non-ergodic at the mode-coupling temperature $T_{\rm MCT}$, distinct from both $T_{\rm SF}$ and $T_{\rm onset}$, showing that even at mean-field level free-energy and energy landscapes are different objects. 

In numerical studies, the relaxation dynamics in $D = 2$ and $3$ ($D$ is the space dimension) harmonic spheres just above jamming was recently studied starting from high temperatures including random configurations at $T=\infty$~\cite{Chacko2019}, and a power-law time decay was found with a non-trivial, dimension-dependent exponent. Another recent work~\cite{Gonzalez-Lopez2020} explores the statistics of single particle displacements between initial and final configurations in several \rev{three-dimensional} models and reports the existence of a crossover temperature separating high- from low-temperature behaviours. \rev{These interesting studies do not provide a complete physical picture of the relaxation dynamics towards energy minima, neither do they assess the existence of the critical temperature $T_{\rm SF}$ found in mean-field approaches. The universality of the power-law time dependence found near jamming across models, and even the effect of spatial dimension and initial temperatures were not fully elucidated, either.}

\rev{Here, we provide a comprehensive numerical study of the steepest descent dynamics in generic glass-forming liquids.} We address its dimensionality, universality, and initial stability dependences by studying a mean-field Mari-Kurchan model and three finite-dimensional models in two, three, four, and eight dimensions using a wide range of initial states obtained through the swap Monte-Carlo algorithm. We numerically detect the predicted mean-field transition at $T_{\rm SF}$ in the Mari-Kurchan model. However, the transition is absent in all finite dimensional models, where it is replaced by a smooth temperature evolution between two \rev{non-trivial limits} that we analyse in detail. We show that this crossover is controlled by a finite population of localised defects where particle rearrangements take place during the minimisation, \rev{with the overall concentration of these defects decreasing at lower} temperatures. \rev{Therefore, finite dimensional glass-forming systems at finite temperatures cannot be seen as inherent structures excited by small thermal fluctuations, since they are neither described by mean-field energy landscapes nor by a simple picture of non-interacting localised defects. Our results provide a complete picture of the relaxation dynamics in glassy landscapes, and illuminate the role, nature and interactions of localised defects in finite-dimensional structural glasses.}

\section{Results}

\subsection{Steepest descent dynamics}

We numerically solve the equations of motion of steepest descent dynamics,
\begin{equation}
\label{eq:SD}
\zeta\frac{\mathrm{d} \vec r_i}{\mathrm{d} t} = - \frac{\partial E}{\partial \vec r_i},
\end{equation}
starting at time $t=0$ from an equilibrium configuration prepared at \rev{initial temperature $T$}, where $\zeta$ is the damping coefficient and $E$ is the potential energy. The time unit is $\tau_0 = \zeta \ell^2 / v_0$, where $\ell$ is the unit length scale, and $v_0$ is the unit energy for particle interactions. \rev{In \eq{eq:SD}, energy is dissipated via a uniform background. We have not tested more complicated dissipation mechanisms such as used in dense particle suspensions~\cite{olsson2015relaxation}. Note that the dynamics in Eq.~(\ref{eq:SD}) is fully athermal (there is no noise term) and the temperature $T$ that we vary only controls the Boltzmann distribution from which initial conditions for the dynamics are drawn.}

We monitor the mean energy $\langle E(t) \rangle$ and the \rev{root mean squared velocity,} 
\begin{equation}
\langle |\vec v (t)| \rangle = \bigang{\sqrt{\frac1N \sum_i \left|\frac{\mathrm{d} \vec r_i}{\mathrm{d} t}\right|^2}},
\end{equation}
during the relaxation dynamics, where the brackets represent an average over initial equilibrium configurations, and $N$ is the number of particles. 
We define an exponent $\beta$ for the time decay~\cite{Chacko2019} as
\begin{align}
  \langle |\vec v (t)| \rangle \sim t^{-\beta}.
  \label{eq:beta}
\end{align}
For the dynamics in \eq{eq:SD}, the energy decay is exactly related to the velocity decay as $\frac{1}{N} \frac{\mathrm{d}}{\mathrm{d}t} \langle E(t) \rangle = - \zeta \langle |\vec v (t)|^2 \rangle$. 
As a result, the energy decay can be expressed using the same exponent: 
$\langle E(t) - E(t \to \infty) \rangle \sim t^{-(2 \beta-1)}$. Therefore, we  focus on the velocity relaxation and Eq.~(\ref{eq:beta}).

We consider several structural glass models in various dimensions and interaction potentials over a wide range of preparation temperatures. We study a soft sphere version of the mean-field Mari-Kurchan model~\cite{Mari2011}, polydisperse soft sphere models in two~\cite{berthier2019zero} and three dimensions~\cite{Ninarello2017}, harmonic spheres~\cite{berthier2009glass} in two, three, four, and eight dimensions, and the Kob-Andersen model~\cite{Kob1995} for two and three dimensions. Note that soft sphere models have a steep repulsive interaction with an $r^{-12}$ core and a short cutoff \revv{(we have checked that extremely few rattler particles~\cite{C7SM01976A} are found in the corresponding inherent structures),} whereas the harmonic potential models have a very soft core, which may affect the $T \to \infty$ limit for initial conditions. The Kob-Andersen model uses the Lennard-Jones potential with a steep repulsive core and attractive forces at larger distances. 

To prepare equilibrium configurations in a wide range of temperatures, we use the planting method~\cite{krzakala2009hiding} for the Mari-Kurchan model and the swap Monte-Carlo algorithm~\cite{Ninarello2017} for some of the finite-dimensional systems, which should allow us to  detect any of the putative transitions predicted from mean-field landscapes.
Further details about the models and simulation protocols are provided in \App{methods} and Supplemental Material \cite{Supplement}.

\subsection{Mean-field Mari-Kurchan model}

\begin{figure}[tbp]
\includegraphics[width=\linewidth]{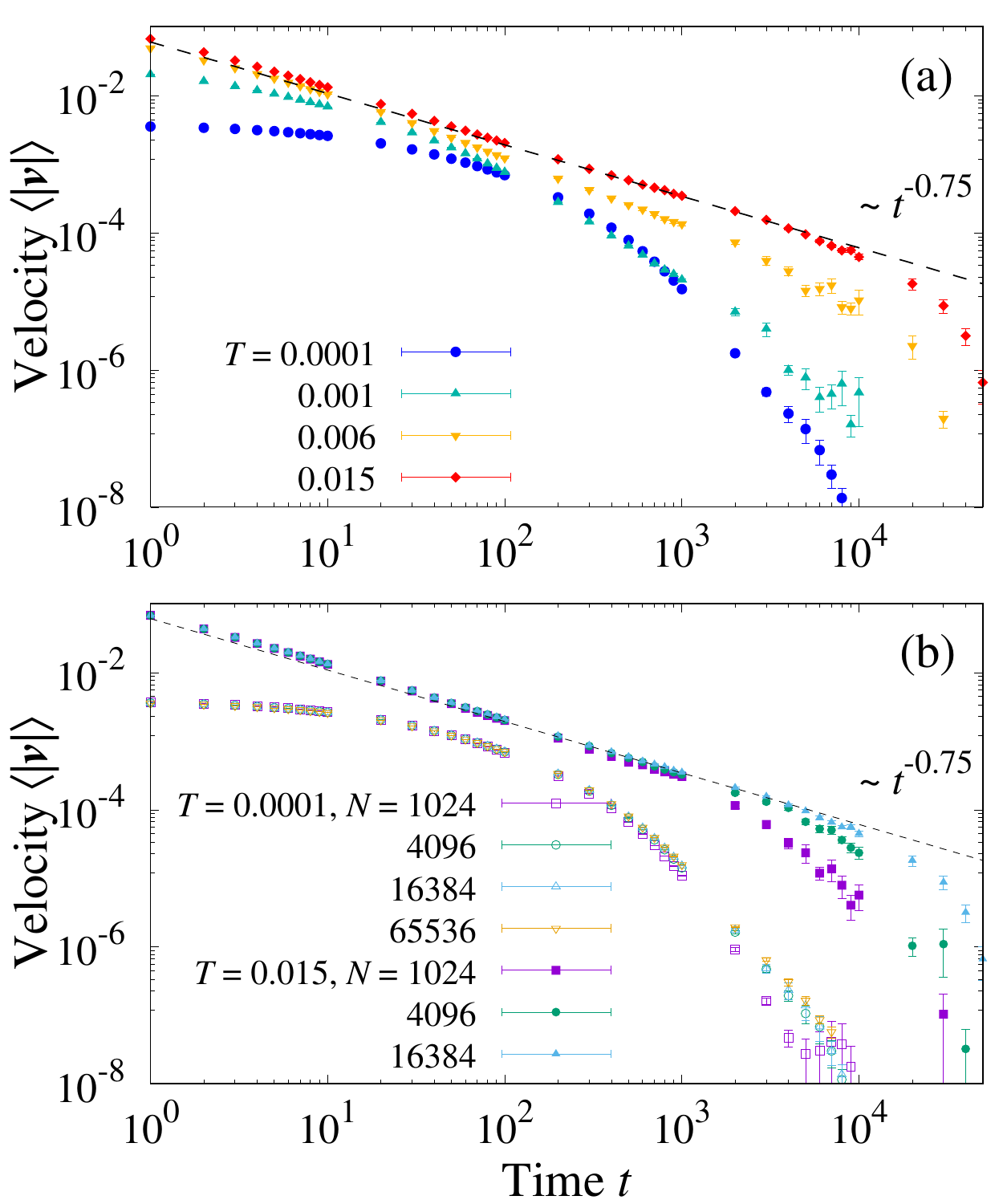}
\caption{
(a) Velocity $\langle |\vec v(t)| \rangle$ as a function of time $t$ in the Mari-Kurchan soft-sphere model at several temperatures with $N = 16384$. (b) Velocity decay at two selected initial equilibrium temperatures and several system sizes. 
The dashed line indicates $\beta=0.75$.}
\label{fig:MK}
\end{figure}

Thanks to its mean-field nature, we can apply the replica liquid theory to the Mari-Kurchan model, as detailed in SI, and obtain the dynamical mode-coupling transition temperature: $T_\text{MCT} \simeq 0.0084$. We also studied the equilibrium dynamics using a simple Metropolis algorithm, and find that the theoretical estimate of $T_\text{MCT}$ describes the numerical data reasonably well. This study allows us to also estimate the onset temperature for slow dynamics: $T_\text{onset} \simeq 0.015$. 

We study the steepest descent starting from equilibrium configurations in the range  $T \in [0.0001, 0.015]$. Figure~\ref{fig:MK} shows the velocity decay $\bigang{|\vec v(t)|}$ for various temperatures and system sizes. Figure~\ref{fig:MK}(a) shows that the relaxation dynamics strongly depends on \rev{initial equilibrium temperature.} For high temperatures, $T \gtrsim 0.006$, $\bigang{|\vec v(t)|}$ follows a clear power-law decay with an exponent that we estimate as $\beta \simeq 0.75$. In a finite size system, this power law decay is interrupted at long times.  On the other hand, at low temperatures, an exponential decay occurs. These results suggest that the high and low temperature relaxation dynamics are qualitatively different, and are separated by a critical temperature. 

To fully confirm the distinct occurrence of power-law and exponential decays, we analyse finite-size effects. In \Fig{fig:MK}(b), we show the velocity $\bigang{|\vec v(t)|}$ for several system sizes at two selected temperatures. At high \rev{initial} temperature $T = 0.015$, $\bigang{|\vec v|}$ has a strong system-size dependence. Larger systems take longer times to reach energy minima and follow the power-law decay with $\beta \simeq 0.75$ over a broader time window. The system-size dependence suggests that in the thermodynamic limit, $N \to \infty$, the velocity decay has a genuine power-law behaviour with a diverging time scale. At very low \rev{initial} temperature, $T = 0.0001$ instead, the velocity decay has almost no system-size dependence, implying that the time to reach energy minima remains finite in the thermodynamic limit, confirming the exponential decay. 

\begin{figure*}[t]
\includegraphics[width=\linewidth]{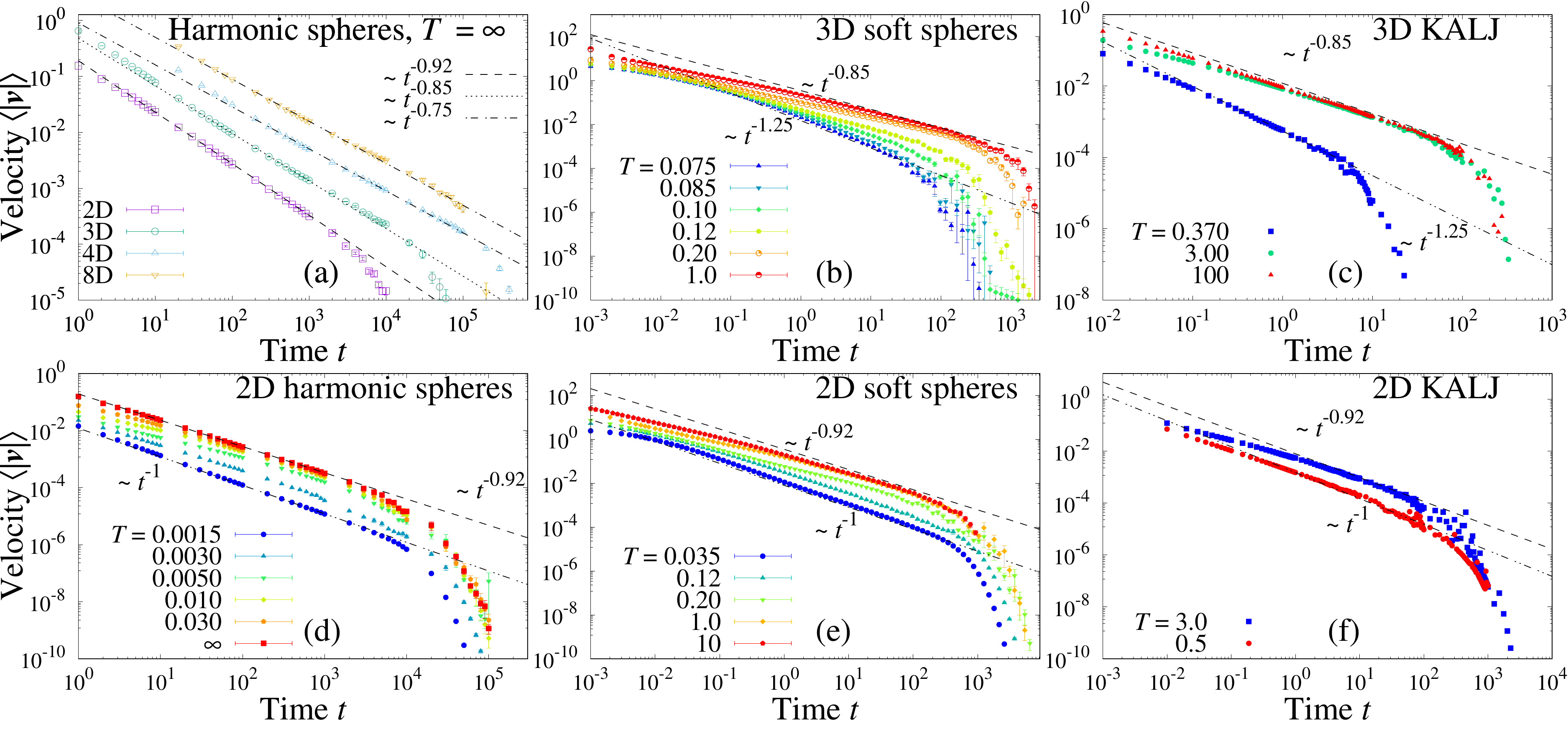}
\caption{
The velocity $\langle |\vec v(t)| \rangle$ as a function of time $t$ for the harmonic spheres in two, three, four, and eight dimensions (a), the three-dimensional soft-spheres (b), the Kob-Andersen Lennard-Jones model (c), the two-dimensional harmonic spheres (d), the soft spheres (e), and the Kob-Andersen Lennard-Jones model (f).
In (a), the number of particles is $N=64000$ for $D=2$, $N=65536$ for $D=3$ and $D=4$, and $N = 16384$ for $D=8$ (data are shifted vertically for clarity).
In (b), (d), (e), and (f), $N=96000$, $64000$, $64000$, and $125000$, respectively. In (c), $N = 27135$ at $T = 0.37$ and $N = 76800$ for other temperatures. }
\label{fig:2348d}
\end{figure*}

Therefore, in the \rev{initial} temperature regime between $0.0001$ and $0.006$, the Mari-Kurchan model displays a transition characterising the nature of the relaxation dynamics akin to the behaviour reported at the temperature $T_{\rm SF}$ discussed in $p$-spin models. Interestingly, for the MK model, the $T_\text{SF}$ is noticeably smaller than the estimated $T_\text{MCT} \approx 0.0084$, confirming that these two temperatures should be distinguished (they are equal in some versions of $p$-spin models).  

While the existence of the transition is compatible with results for the mixed $p$-spin glass model, the decay exponent $\beta \simeq 0.75$ at high temperatures observed for the Mari-Kurchan model differs slightly from the spin glass model where $\beta \simeq 0.83$  \footnote{The exponent $\beta$ for the spin glass model is obtained from the exponent for the energy decay reported in \cite{Folena2020} which is given by $2\beta -1$.}, \revv{or the random Lorentz gas in $d \to \infty$ where $\beta \approx 1$~\cite{manacorda2022gradient}.} A broader class of mean-field models and even more extensive numerics should be explored to settle the relevance of this small difference. 

\subsection{Finite-dimensional models}

\label{sec:finited_velocity}

For finite $D$ systems, we first consider the relaxation dynamics starting from the high-temperature limit, $T \to \infty$, in various models and spatial dimensions. Figure~\ref{fig:2348d}(a) shows the results for monodisperse harmonic spheres in dimensions $D = 2$, 3, 4, and 8. In all dimensions, we observe a power-law decay, but the exponent $\beta$ depends on $D$. We find $\beta\simeq 0.92$ for $D=2$ and $\beta \simeq 0.85$ for $D=3$, which are consistent with previous work~\cite{Chacko2019}. For $D=4$ and 8, on the other hand, the results are described by the same exponent $\beta \simeq 0.75$. This suggests that a mean-field value $\beta=0.75$ in the high temperature regime is reached for $D \geq 4$. Interestingly, this exponent is close to the one observed in the Mari-Kurchan model at high temperatures. In SI, we discuss the system-size dependence of $\bigang{|\vec v(t)|}$. Larger systems always take a longer time to reach energy minima, consistently with a pure power law decay in the thermodynamic limit $N \to \infty$ at large times.

To investigate universality, we look at the results at high temperatures for various models. Figures~\ref{fig:2348d}(b-f) show the velocity decay for harmonic and soft spheres, and the Kob-Andersen model. Note that for soft spheres and Kob-Andersen models, the influence of the repulsive core can be felt at arbitrarily large temperatures. Nevertheless, all models at higher \rev{initial} temperatures asymptotically show $\beta \simeq 0.92$ and $0.85$ in $D=2$ and $D=3$, respectively. Therefore, we conclude that the value of $\beta$ is universal, irrespective of the details of the interaction potentials, size polydispersity, or the proximity of a jamming transition.

We then study the effect of the initial stability on the relaxation dynamics. We first consider $D=3$ polydisperse soft spheres. Using the swap Monte Carlo algorithm, we vary \rev{initial equilibrium temperature} quite significantly, $T \in [0.062, 1.0]$, which includes $T_\text{MCT} \simeq 0.1$ and $T_\text{onset} \simeq 0.18$ determined by standard methods~\cite{Ninarello2017}. The model reproduces the universal exponent in $D=3$, $\beta=0.85$, at finite but high temperatures, see \Fig{fig:2348d}(b). With decreasing temperature, the velocity relaxation $\bigang{|\vec v(t)|}$ becomes faster, as expected from the physical intuition that the system starts closer to an energy minimum in a smoother landscape. However, even at temperatures much below $T_\text{MCT}$, the velocity relaxation $\bigang{|\vec v(t)|}$ displays a power-law decay, but with a larger apparent exponent, $\beta \simeq 1.25$. The same exponent at low $T$ is found in the $D=3$ Kob-Andersen model [see \Fig{fig:2348d}(c)], suggesting that $\beta \simeq 1.25$ is also universal. We vary the initial stability for $D=2$ harmonic spheres, soft spheres, and the Kob-Andersen model in \Fig{fig:2348d}(d), (e), and (f). The data demonstrate the same trend as in $D=3$ models, yet the low-temperature velocity decay exponent is now $\beta \simeq 1$ in all three models, different from the $D=3$ value. Importantly, we do not observe an exponential decay at any studied temperatures in any of the finite-dimensional models, in contrast to the mean-field spin glass and Mari-Kurchan models. Instead, we find that the high- and low-temperature regimes are both characterised by universal power-laws, with an exponent which only depends on the spatial dimension. 

\subsection{Harmonic limit}

\label{sec:harmonic}

We can rationalize our numerical observations at low temperatures using a harmonic dynamical description \footnote{The term `harmonic' here means that the system energy is expanded up to the quadratic term. This harmonic approximation is applicable to any smooth interaction potential, and not only to the system of harmonic spheres.} At very low \rev{initial} temperatures, the initial equilibrium configuration is located nearer to the final inherent state. Thus, it makes sense to approximate the energy during the steepest descent dynamics using \rev{a harmonic expansion}
\begin{equation}
E(t) \simeq E(t \to \infty) + \frac12 \Delta \vec r(t) \cdot H \cdot \Delta  \vec r(t),
\end{equation}
with $\Delta \vec r(t) = \vec r(t \to \infty) - \vec r(t)$ and $H$ the Hessian matrix in the energy minimum. Let us assume that the phononic modes following the Debye law and quasi-localised modes following the non-Debye quartic law coexist in the low-frequency region of the vibrational density of states~\cite{Mizuno2017}. \rev{By linearising the equations of motion, we can relate the time decay of the velocity to the properties of the Hessian matrix} and we find that the velocity should decay with an exponent $\beta_\text{harm} = D/4 + 1/2$, yielding $\beta_\text{harm} = 1$ and 1.25 for $D=2$ and $D=3$, respectively. (See \App{appendix:harmonic} for a more detailed discussion of the harmonic approximation, which also shows that quasi-localised modes provide a subdominant contribution to the velocity decay for $D < 5$.) These values are fully consistent with our numerical observations, which means that the low-temperature relaxation dynamics appears to be well-described, at least over the simulated timescales and system sizes, by a simple harmonic approach, \rev{for all types of particle interactions.}

\subsection{Localised defects}

The harmonic analysis shows that, because of phonons, the mean-field transition at $T_{\rm SF}$ to gapped energy minima reached exponentially fast cannot exist in finite $D$, and relaxation dynamics is in fact necessarily algebraic, even in a harmonic, `state following' limit. Our numerics is nevertheless compatible with two distinct temperature regimes, with non-harmonic effects becoming predominant at high \rev{initial} $T$. Is a sharp transition separating these two regimes? 

\begin{figure}[t]
\includegraphics[width=\linewidth]{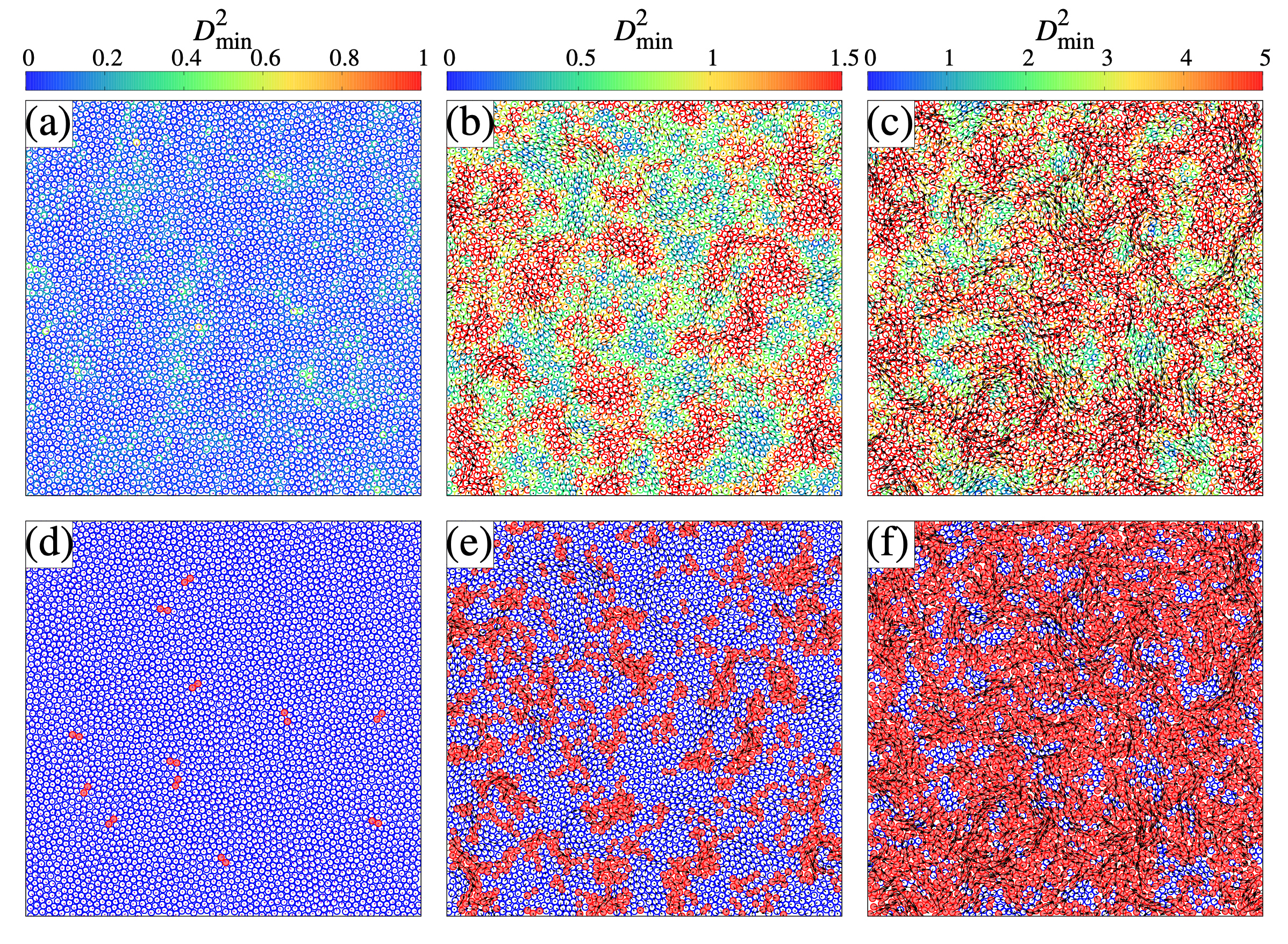}
\caption{$2D$ soft spheres.
  (a-c) Non-affine displacements $D_\text{min}^2$ and (d-f) defects for three different configurations. In (d-f), defects are shown in red, other particles  are in blue. In all panels, the displacement vectors $\vec r_i(t=0) - \vec r_i(t \to \infty)$ (arrows) is amplified by a factor of $3$. The temperature is $T=0.035$ in (a, d), $0.2$ in (b, e), and $0.8$ in (c, f).}
\label{fig:defect_snapshots}
\end{figure}

To address this question, we must understand the microscopic relaxation mechanism beyond the harmonic limit. At very low temperatures, we expect that the initial configuration and the final inherent state differ by small displacements which do not affect much the geometry of the particle packing. Figure~\ref{fig:defect_snapshots}(a-c) show the displacement and non-affine displacement fields~\cite{Falk1998} between initial and final configurations for the $D=2$ soft sphere system. For $T=0.035$, where the harmonic description works well (and $\beta = \beta_{\rm harm}=1.0$ is measured), particle displacements are indeed very small, implying that most particles interact with the same neighbours in initial and final states, see \Fig{fig:defect_snapshots}(a). For $T=0.2$, however, larger displacements are observed, and the most mobile particles are spatially correlated, see \Fig{fig:defect_snapshots}(b). Large non-affine displacements are associated with localised particle rearrangements occurring during the steepest descent, which we call `defects'. As the temperature is increased further, more particles have large non-affine displacements, see \Fig{fig:defect_snapshots}(c), and the initial and final configurations become substantially different.

To quantify particle rearrangements during minimisation, we introduce a variable $\phi_i$ for each particle defined such that $\phi_i=0$ if particle $i$ neither loses nor gains any neighbour during steepest descent, and $\phi_i=1$ otherwise (see SI for precise definitions), and denote $\phi = \frac{1}{N} \sum_i \phi_i$ the concentration of such defects in a given configuration with $N$ particles. The field $\phi_i$ thus identifies the location of particle rearrangements, as shown in \Fig{fig:defect_snapshots}(d-f). Red particles with $\phi_i=1$ are found in high $D^2_\text{min}$ regions, which validates the proposed identification of defects. The defects are also observed in the displacement field $|\vec r_i(t) - \vec r_i(\infty)|$, see SI for further discussion. In Ref.~\cite{Chacko2019}, similar defects were visualised using the non-affine velocity field. 

\begin{figure}[t]
\includegraphics[width=\linewidth]{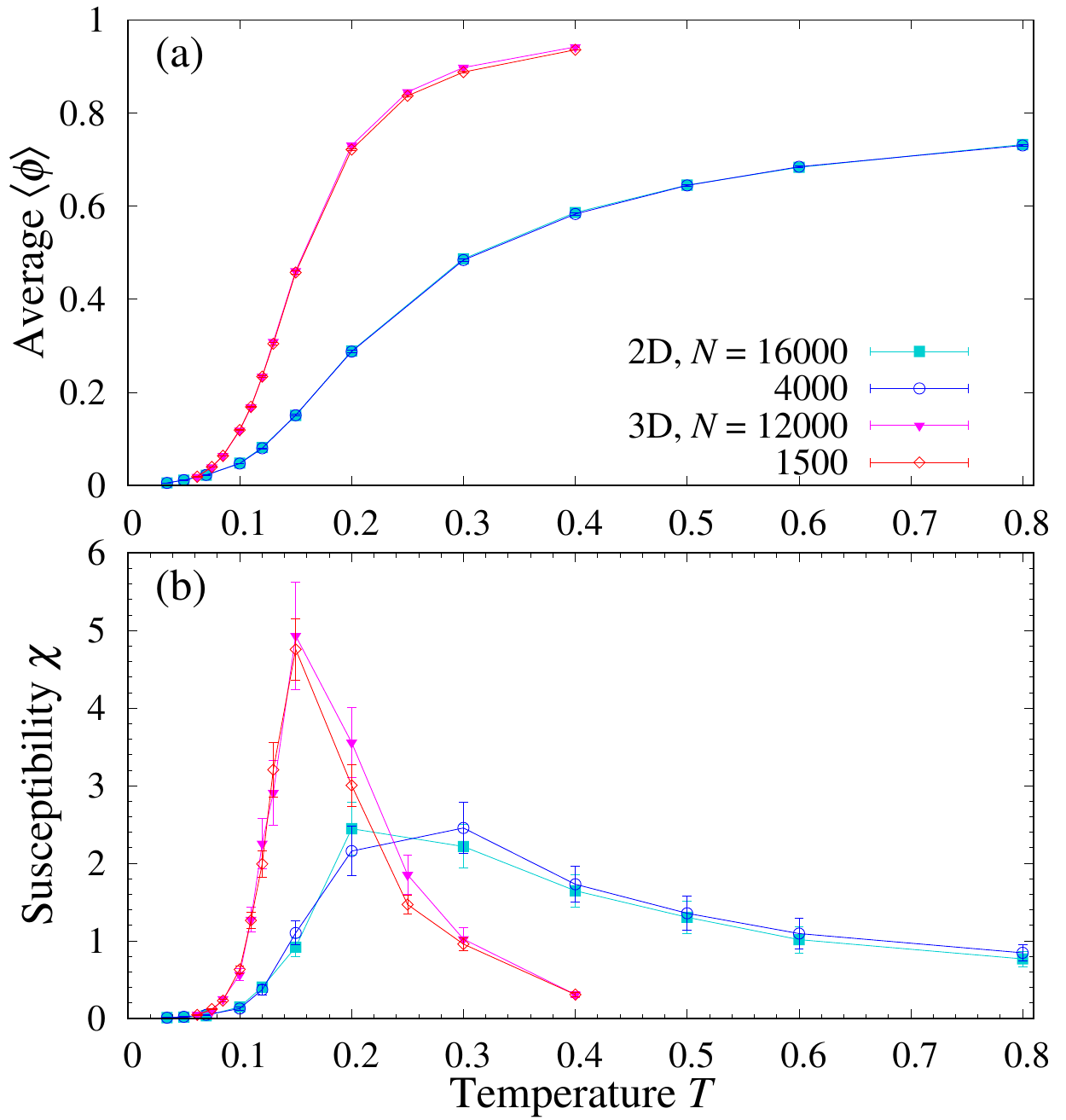}
\caption{Soft spheres. 
(a) Average concentration $\langle \phi \rangle$ and (b) collective susceptibility $\chi$ of defects, as a function of \rev{initial equilibrium} temperature. The temperature evolution of the defect concentration is smooth, with a maximum variation near $T_\text{def} \approx 0.25$ and $T_\text{def} \approx 0.15$ for $D=2$ and $3$, respectively.}
\label{fig:defect}
\end{figure}

In \Fig{fig:defect}, we show the average concentration of defects, $\langle \phi \rangle$, for $D=2$ and $D=3$ soft-sphere models, and the collective susceptibility of defects, $\chi = N( \langle \phi^2 \rangle -  \langle \phi \rangle^2)$. The average defect density is a smooth function of temperature which seems to remain finite at any \rev{initial} $T>0$. The susceptibility shows a well-defined peak, whose shape and location are independent of the system size, see Fig.~\ref{fig:defect}(b). These results indicate that no sharp phase transition (with a vanishing $\langle \phi \rangle$) separates the relaxation dynamics between high and low temperatures. The defect density has a sigmoidal shape as it saturates to unity at large $T$ and decreases very rapidly to small values as $T \to 0$. It displays an inflection point at a temperature $T_\text{def}$ that also corresponds to the peak of the susceptibility. Physically, $T_{\rm def}$ represents the temperature where $\langle \phi \rangle$ varies more strongly with $T$ and has the largest fluctuations, thus separating the high-$T$ regime where $\langle \phi \rangle$ approaches unity, from low-$T$ where it is very small. The gradual disappearance of localised rearrangements \rev{presumably explains the} temperature evolution of the self-part of the van-Hove function~\cite{Gonzalez-Lopez2020}. The discussion of the harmonic limit in Sec.~\ref{sec:harmonic} showed that the defects revealed by steepest descent dynamics at lower temperatures do not simply result from the harmonic excitation of the quasi-localised modes populating the low-frequency part of the density of states (which would lead to a different power law decay), although a more complicated relation could exist.

\section{Discussion}

We studied the physical dynamics during steepest descent energy minimisation for various glass-forming models in spatial dimensions $D=2$ to $D=8$ and also in the mean-field limit, for a wide range of initial conditions. Focusing on the exponent $\beta$ characterising the algebraic decay of the average velocity, we identified its universal, finite dimensional features. First, we showed that the mean-field transition at temperature $T_{\rm SF}$ to an exponential decay cannot exist in finite $D$ due to the presence of phonons. More importantly, we showed that the measured evolution of $\beta$ from its high-temperature universal value towards a larger harmonic value $\beta_{\rm harm} = D/4+1/2$, observed at low-temperatures, reflects in fact the gradual suppression of a population of localised defects with decreasing $T$. The relative importance of defects and plane waves explains the observed evolution of $\beta$. Since $\beta_{\rm harm}$ is larger than its high-$T$ value, we expect the latter exponent to dominate the long-time limit of the velocity decay at any finite temperature in the thermodynamic limit. In this view, the harmonic regime is only a transient which lasts longer at lower temperature when there are less defects. As a result, the mean-field critical temperature $T_{\rm SF}$ has no analog in finite $D$. This implies that, at finite temperature, an instantaneous configuration of a finite dimensional glass-forming system \rev{can never be seen} as an inherent structure excited by small thermal fluctuations. It would be interesting to explore theoretical models alternative to mean-field glass models, such as elasto-plastic models~\cite{parley2020aging}, to account better for our numerical observations, in particular the value of the exponent $\beta$.

Our results have broad physical consequences. First, they imply that the defect dynamics leading to the coarsening of the non-affine velocity field described in Ref.~\cite{Chacko2019} (see also SI) is actually relevant for generic finite $D$ glass-forming liquids, and is unrelated to the athermal jamming transition. The observed universal exponent $\beta$ implies similarly universal geometrical features of the potential energy landscapes of generic structural glasses. \rev{Interestingly, an experimental realisation of the steepest descent dynamics has recently been proposed~\cite{Yanagisawa2021}. By perturbing a stable foam configuration in two dimensions, localised defects during the relaxation were also observed. Such experiments could validate our numerical findings, especially the universal exponent $\beta$ found at high initial temperatures. More generally, our observations about defects at lower initial temperatures is another supporting evidence of the existence of localised excitations in stable glasses relevant for metallic and molecular glasses~\cite{qiao2019structural}.} 

Second, together with recent analytic and numerical works~\cite{Folena2020,Coslovich2019}, our results shed new light on the connection between equilibrium glassy dynamics and stationary points of the potential energy landscape. The interpretation of the mode-coupling temperature $T_{\rm MCT}$ as a topographic change in the potential energy landscape does not hold in mixed $p$-spin models~\cite{Folena2020}. Our simulations of the Mari-Kurchan model confirm that the saddle-to-minima transition occurs at a temperature $T_{\rm SF}$ distinct from $T_{\rm MCT}$, already at mean-field level. The emergence of localised defects in finite dimensions found here is consistent with the recent conclusion~\cite{Coslovich2019} that the critical transition at $T_{\rm SF}$ is replaced in finite $D$ by a smooth crossover. A similar scenario controlled by non-interacting localised defects was also found in kinetically constrained models~\cite{berthier2003real}, thus suggesting a potential connection between the defects revealed by steepest descent dynamics and those discussed in the context of dynamic facilitation~\cite{keys2011excitations}. \rev{However, the power law decay revealed by our study cannot result from the de-excitation of a non-interacting gas of isolated defects and steepest descent dynamics in kinetically constrained models would instead be unremarkable. It is also unclear whether elasto-plastic models where relaxation events are coupled by elasticity can account for our findings.}  

\begin{figure}[t]
\includegraphics[width=\linewidth]{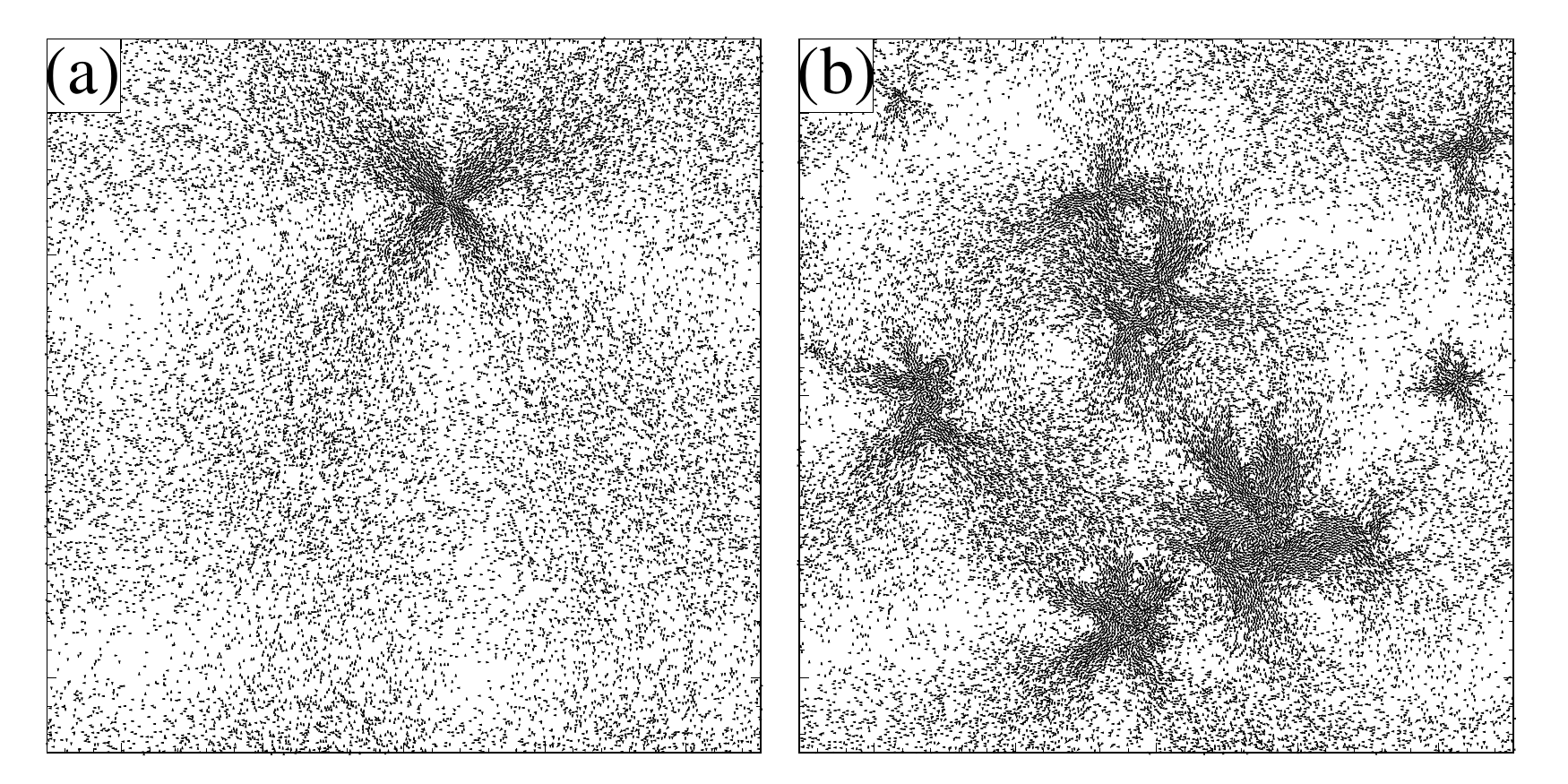}
\caption{Displacement fields, $\vec{r}_i^{\rm CG}-\vec{r}_i^{\rm FIRE}$, between pairs of minima obtained via conjugate gradient ($\{ \vec{r}_i^{\rm CG} \}$) and FIRE algorithms ($\{ \vec{r}_i^{\rm FIRE} \}$), following a quench from the same initial configuration at $T=0.07$ (a) and $T=0.15$ (b), for $2D$ soft spheres with $N=64000$. Arrows magnified by a factor $40$ and $2$ for (a) and (b), respectively.}
\label{fig:disp_cgfire}
\end{figure}

Third, our finding that a finite concentration of defects controls the non-harmonic relaxation from equilibrated configurations to inherent states suggests that the potential energy landscape of glass-formers is both rugged and chaotic. To test this idea numerically, we applied a very small random perturbation to the initial configuration and monitored the subsequent steepest descent dynamics. We found that a slight perturbation typically leads to different inherent structures (not shown), consistent with earlier work~\cite{scalliet2017absence,Scalliet2019}. The strong chaoticity of the minimisation dynamics implies that the energy minimum reached from a given equilibrium configuration in fact strongly depends on the minimisation algorithm itself~\cite{angelani2003general}. The steepest descent (SD) dynamics we used is just the simplest algorithm for numerical optimisation, but there are several other (usually more efficient) ways to reach the bottom of the potential energy landscape, such as conjugate-gradient (CG)~\cite{nocedal2006numerical} and fast inertial relaxation engine (FIRE)~\cite{Bitzek2006}. Indeed, we find that starting from the same initial configuration, the SD, CG and FIRE dynamics typically converge to different inherent structures, as quantified by their mutual distances (see SI). The evolution of this distance mirrors the temperature evolution of the defect concentration in Fig.~\ref{fig:defect}(a), higher initial temperatures leading to larger separations. In \Fig{fig:disp_cgfire}, we show representative snapshots of the displacement field between two inherent structures obtained by two different algorithms starting from a unique initial configuration. A single localised defects can be seen at low $T$, which naturally gives rise to a quadrupolar Eshelby-like displacement field. Defects proliferate at higher temperature. This shows that mapping an equilibrium liquid state to an inherent structure is a fully dynamical problem, which becomes uniquely defined only after a specific choice for the minimisation algorithm is made. Localised defects, which had been used by Stillinger~\cite{stillinger1988supercooled} to construct an argument against the existence of a Kauzmann transition (see the discussion in \cite{berthier2019configurational}), instead weaken the thermodynamic significance of a tiling of configuration space directly based on inherent states.

In recent years, localised glassy defects have been reported from the study of harmonic~\cite{lerner2016statistics} and non-harmonic excitations, in the fields of plasticity~\cite{richard2020predicting} and low-temperature transport properties~\cite{khomenko2020depletion}, and in connection with secondary relaxations in deeply supercooled liquids~\cite{yu2017structural,guiselin2021microscopic} and dynamic facilitation~\cite{keys2011excitations}. \rev{Steepest descent dynamics thus corresponds to another situation where localised excitations control structural rearrangements at the particle scale and reveal that they interact in a non-trivial manner. Future work should establish the similarities and differences between these disparate observations. Ultimately, we expect that a unifying picture of localised defects with specific interactions will soon become available and applicable to a host of different physical situations.} 

\begin{acknowledgments}
We thank G. Biroli, R. Chacko, G. Folena, and H. Ikeda for discussions. We also thank A. D. S. Parmar for sharing stable Kob-Andersen configurations. This work was supported by grants from the Simons Foundation (\#454935 L. Berthier) and JSPS KAKENHI (Grants No. 18H05225, 19H01812, 20H01868, 20H00128, A. Ikeda). 
\end{acknowledgments}

\appendix
\section{Methods}
\label{methods}
\subsection{Models}
We study the steepest descent dynamics of models with three different interaction potentials: soft spheres, harmonic spheres, and Lennard-Jones interactions. The dimensionality dependence, including the mean-field limit, of this dynamics is studied by using the models in two-, three-, four-, eight-dimensions, and the mean-field Mari-Kurchan model~\cite{Kraichnan1962,Mari2011}.

\subsubsection{Soft spheres}

The two- and three-dimensional soft sphere models~\cite{berthier2019zero,Ninarello2017} consist of particles with purely repulsive interactions and a continuous size polydispersity. Particle diameters, $d_i$, are randomly drawn from a distribution of the form: $f(d) = Ad^{-3}$, for $d \in [ d_\text{min}, d_\text{max} ]$, where $A$ is a normalization constant. The size polydispersity is quantified by $\delta=(\overline{d^2} - \overline{d}^2)^{1/2}/\overline{d}$, where the overline denotes an average over the distribution $f(d)$. Here we choose  $\delta = 0.23$ by imposing $d_\text{min} / d_\text{max} = 0.449$. The average diameter, $\overline{d}$, sets the unit of length. The soft-sphere interactions are pairwise and described by an inverse power-law potential
\begin{align}
\label{eq:ss}
&U_{ij}(r) = v_0 \left( \frac{d_{ij}}{r} \right)^{12} + c_0 + c_1 \left( \frac{r}{d_{ij}} \right)^2 + c_2 \left( \frac{r}{d_{ij}} \right)^4,\\
&d_{ij} = \frac{(d_i + d_j)}{2} (1-\epsilon |d_i - d_j|), \nonumber
\end{align}
where $v_0$ sets the unit of energy (and of temperature with the Boltzmann constant $k_\mathrm{B}\equiv 1$) and $\epsilon=0.2$  quantifies the degree of nonadditivity of particle diameters. We introduce $\epsilon>0$ in the model to suppress fractionation and thus to enhance the glass-forming ability. The constants $c_0$, $c_1$ and $c_2$ enforce a vanishing potential and continuity of its first- and second-order derivatives at the cut-off distance $r_\text{cutoff}=1.25 d_{ij}$.
We simulate a system with $N$ particles within a square cell of area (volume) $V=L^2$ ($V=L^3$) where $L$ is the linear box length, under periodic boundary conditions, at a number density $\rho=N/V=1$ (1.02) for $2D$ ($3D$).

We prepare equilibrium configurations using the swap Monte Carlo algorithm~\cite{Ninarello2017}. With probability $P_\text{swap}=0.2$, we perform a swap move where we randomly pick two particles ($i$ and $j$) having similar diameters ($|d_i-d_j|<0.2$) and attempt to exchange their diameters. With probability $1-P_\text{swap}=0.8$, instead, we perform conventional Monte Carlo translational moves, where we pick one particle and displace it within a box with linear length $\delta_\text{max}=0.12 \overline d$.

\subsubsection{Harmonic spheres}

We study the harmonic sphere model~\cite{o2003jamming,berthier2009glass} in two, three, four, and eight dimensions.
The harmonic sphere model has an interaction potential
\begin{align}
&U_{ij}(r_{ij}) = \frac{v_0}2 \left( 1 - \frac{r_{ij}}{d_{ij}}\right)^2, \\
&d_{ij} = \frac{(d_i + d_j)}{2} (1-\epsilon |d_i - d_j|),
\end{align}
where $v_0$ is again the unit of the energy scale. For the two dimensional model, to avoid crystallisation at low temperature, we use the continuously polydisperse non-additive model with the same distribution of the particle diameters used in the soft-sphere model and $\epsilon=0.2$ in two dimensions. The unit length scale for the two-dimensional model is $\overline{d}$ as well as the soft-sphere model. We again use the swap Monte Carlo algorithm with the same setting and parameters as for the polydisperse soft spheres to equilibrate down to very low temperatures. In three, four, and eight dimensions, crystallisation is highly suppressed, and the simple additive ($\epsilon=0$) monodisperse model is enough to study the relaxation dynamics to disordered states. Due to the finite range of the interaction, the system has a critical jamming transition at finite density, below which the relaxation dynamics shows an exponentially fast decay towards zero energy states~\cite{Nishikawa2021}. Since in other models we study the relaxation dynamics towards energy minima with a finite energy, a direct comparison is possible when the inherent structures of harmonic spheres have finite energies as well. We thus set the volume fraction above the jamming transition to $\phi = 1.2$, $0.73$, $0.5$, $0.1$ in two, three, four, and eight dimensions, respectively, so that the final energies are always finite.

\subsubsection{Kob-Andersen Lennard-Jones model}

For the case of the well-studied Kob-Andersen binary Lennard-Jones (KALJ) model, the interaction between two particles has the following form:
\begin{equation}
U_{ij}(r_{ij}) = 4v_0 \bigpar{ \bigpar{\frac{\sigma_{ij}}{r_{ij}}}^{12} -  \bigpar{\frac{\sigma_{ij}}{r_{ij}}}^6}.
\end{equation}
for $ r < r_\text{cutoff} = 2.5\sigma_{ij}$  and particles $i,j$ can belong to either A or B species which constitute the binary mixture. $r_\text{cutoff}$ is the cutoff distance at which the  potential $U_{ij}(r_{ij})$ is truncated.
The different interaction parameters for the binary mixture take the following values: $\epsilon_{\textrm{AA}} = 1.0$, $\epsilon_{\textrm{AB}} = 1.5\epsilon_{\textrm{AA}}$,
$\epsilon_{\textrm{BB}} = 0.5\epsilon_{\textrm{AA}}$; $\sigma_{\textrm{AA}} = 1.0$,
$\sigma_{\textrm{AB}} = 0.8\sigma_{\textrm{AA}}$, $\sigma_{\textrm{BB}} = 0.88\sigma_{\textrm{AA}}$. The mixture has 80:20 composition in $3D$ and 65:35 composition in $2D$, to optimise glass-forming ability. In $D=2$, we study a system consisting 125000 particles and for $D=3$, we study a system of 76800 particles. Additionally, in $D=3$, we study a system of 27135 particles to probe the quench dynamics of states sampled at a low temperature ($T=0.37$), where configurations are obtained by a swap Monte Carlo scheme developed in Ref.~\cite{parmar2020stable}.

\subsubsection{Mari-Kurchan model}

We study the mean-field Mari-Kurchan (MK) model \cite{Kraichnan1962,Mari2011} in three dimensions with the simple mono-disperse soft-sphere interaction in \eq{eq:ss} with $\epsilon = 0$ and the cutoff length $r_\text{cutoff} = 4d$, where $d$ is the diameter of particles. The volume fraction is $\phi = 0.5$. The MK model has quenched randomness in the particle distance, and the interaction potential is thus $U(|\vec r_i - \vec r_j + \vec A_{ij}|)$, where $\vec A_{ij}$ is a three-dimensional vector with each component sampled from the uniform distribution in the interval $[0, L]$ ($L$ the box size). Equilibrium configurations of the MK model are produced by using the planting technique \cite{Mari2011,Charbonneau2014}. For systems with general isotropic interactions, the cubic shape of the box complicates the direct sampling of the random shifts from the Boltzmann distribution
\begin{equation}
\label{eq:planting}
P(\vec A_{ij}| \vec r_{ij}) 
= \frac{\exp(-\beta U_{ij}(|\vec r_i - \vec r_j + \vec A_{ij}|))}
{\int d\vec A_{ij} \exp(-\beta U_{ij}(|\vec r_i - \vec r_j + \vec A_{ij}|))}.
\end{equation}
We thus use the Markov chain Monte Carlo method to sample the random shifts $\{\vec A_{ij}\}$ from the distribution \eq{eq:planting} so that any given particle configuration follows from the Boltzmann distribution. For each pair of particles $i$ and $j$, we take $\vec A_{ij}$ as the random shift after 200 Monte Carlo sweeps with the simple Metropolis algorithm starting from uniformly random numbers.

\rev{\section{Harmonic exponent}}

\label{appendix:harmonic}

\rev{We discuss the asymptotic decay of the velocity by assuming that the system is perfectly harmonic and the vibrational density of states follows the Debye law. Let the Hessian matrix $H$ of an inherent structure have eigenvalues $\{ \lambda_a\}$ and corresponding eigenvectors $\{\vec x_a\}$. Since the Hessian matrix is real symmetric, eigenvectors are orthogonal; $\vec x_a \cdot \vec x_b = \delta_{ab}$, where $\delta_{ab}$ is the Kronecker's delta. Using the eigenvectors, we have the particle displacement written as
\begin{equation}
\Delta \vec r(t) = \sum_a c_a(t) \vec x_a,
\end{equation}
where $c_a(t) = \Delta \vec r(t) \cdot \vec x_a$. Suppose that the system is perfectly harmonic, i.e. the system follows linearised equations of motion,
\begin{equation}
\zeta\frac{\mathrm{d} \Delta \vec r}{\mathrm{d} t} = - H \cdot \Delta \vec r. 
\end{equation}
Then each mode decays exponentially with $c_a(t) = c_a(0)e^{-\lambda_a t}$ and the equipartition law $\bigang{c_a(0)^2} = \frac{T}{ \lambda_a}$ holds.}

\rev{In this harmonic approximation, the potential energy decreases with time as
\begin{align}
\bigang{E(t) - E(t\to \infty)} 
&= \frac12 \sum_a c_a^2(t) \lambda_a
\exp\bigpar{-2 \lambda_a t} \nonumber \\
&= \frac{NT}{2} \int d\lambda \rho(\lambda) \exp\bigpar{-2\lambda t},
\end{align}
where $\rho(\lambda) = \bigang{\frac1N\sum_a \delta(\lambda-\lambda_a)}$ is the density of eigenvalues. }

\rev{Let us assume that the density of state has the contributions from the phononic modes following the Debye law and quasi-localised modes following the non-Debye quartic law i.e. $g(\omega) = A_0 \omega^{D-1} + A_4 \omega^4$~\cite{Mizuno2017,Kapteijns2018,Shimada2020}. Then the density of eigenvalues reads $\rho(\lambda) = g(\omega) \frac{d\omega}{d\lambda} = A_0 \lambda^{D/2-1} + A_4 \lambda^{3/2}$. Thus
\begin{align}
\bigang{E(t) - E(t\to \infty)} 
&\sim \frac{N T}{2} \int d\lambda (A_0 \lambda^{D/2-1} + A_4 \lambda^{3/2}) 
e^{-2\lambda t} \nonumber \\
& \sim t^{-D/2} + O(t^{-5/2}).
\label{eq:debye}
\end{align} 
Therefore, the energy relaxation is dominated by $t^{-D/2}$ when $D \leq 5$. 
Since, for the steepest descent dynamics with the equations of motion given by Eq.~(1), the energy decay can be related to the velocity decay, we finally obtain $\bigang{|\vec v(t)|} \sim t^{-\beta_\text{harm}}$ with $\beta_\text{harm} = D/4 + 1/2$ for $D \leq 5$. }

\bibliography{refs}

\end{document}

\title{Supplemental Material for ``Relaxation dynamics in the energy landscape of glass-forming liquids''}

\author{Yoshihiko Nishikawa}

\affiliation{Laboratoire Charles Coulomb (L2C), Universit\'e de Montpellier, CNRS, 34095 Montpellier, France}

\author{Misaki Ozawa}

\affiliation{Laboratoire de Physique de l'Ecole normale sup\'erieure, ENS, Universit\'e PSL, CNRS, Sorbonne Universit\'e, Universit\'e Paris-Diderot, Sorbonne Paris Cit\'e, Paris, France}

\author{Atsushi Ikeda}

\affiliation{Graduate School of Arts and Sciences, The University of Tokyo, Tokyo 153-8902, Japan}

\author{Pinaki Chaudhuri}

\affiliation{The Institute of Mathematical Sciences, C.I.T. Campus, Taramani, Chennai 600 113, India}

\author{Ludovic Berthier}

\affiliation{Laboratoire Charles Coulomb (L2C), Universit\'e de Montpellier, CNRS, 34095 Montpellier, France}

\affiliation{Yusuf Hamied Department of Chemistry, University of Cambridge, Lensfield Road, Cambridge CB2 1EW, United Kingdom}

\date{\today}

\maketitle

\section{Replica liquid theory for the Mari-Kurchan (MK) model}

\label{app:rlt}

Thanks to the presence of the random shifts, the replica liquid theory can be applied to the MK model, which enables us to predict the dynamical transition in this model. 
This procedure is well established for the MK model with hard-sphere interactions~\cite{Charbonneau2014,Ikeda2017}, and the extension to soft spheres is straightforward. 
In the replica liquid theory, the cage size $\alpha$ is used as the order parameter, and the self-consistent equation of $\alpha$ for the $3D$ MK model can be written as 
\begin{eqnarray}
\label{eq:mkscf}
\frac{1}{\alpha} = - \frac{2 \phi}{\pi} \int d\vec{r} \frac{\partial Q(\vec{r})}{\partial \alpha} \log Q(\vec{r}), 
\end{eqnarray}
with 
\begin{eqnarray}
Q(\vec{r}) = \int d\vec{r}^\prime \gamma_{2\alpha} (\vec{r} + \vec{r}^\prime) e^{- \beta U(|\vec{r}^\prime|)}, 
\end{eqnarray}
where $\gamma_{\alpha}(\vec{r}) = (2\pi \alpha)^{-3/2} e^{- \frac{|\vec{r}|^2}{2\alpha}}$ is the normalized gaussian function, and $U(r)$ is the soft-sphere interaction potential Eq.(5). 
The dynamical transition temperature $T_\text{MCT}$ is defined as the highest temperature at which the self-consistent equation \eq{eq:mkscf} has a solution. 
We numerically solved this equation and obtained $T_\text{MCT} \simeq 0.0084$ and $\alpha = 0.13$ at $T = T_\text{MCT}$.  

\section{System-size dependence of the velocity decay}

\label{app:Ndep}

\begin{figure}[b]
\includegraphics[width=\linewidth]{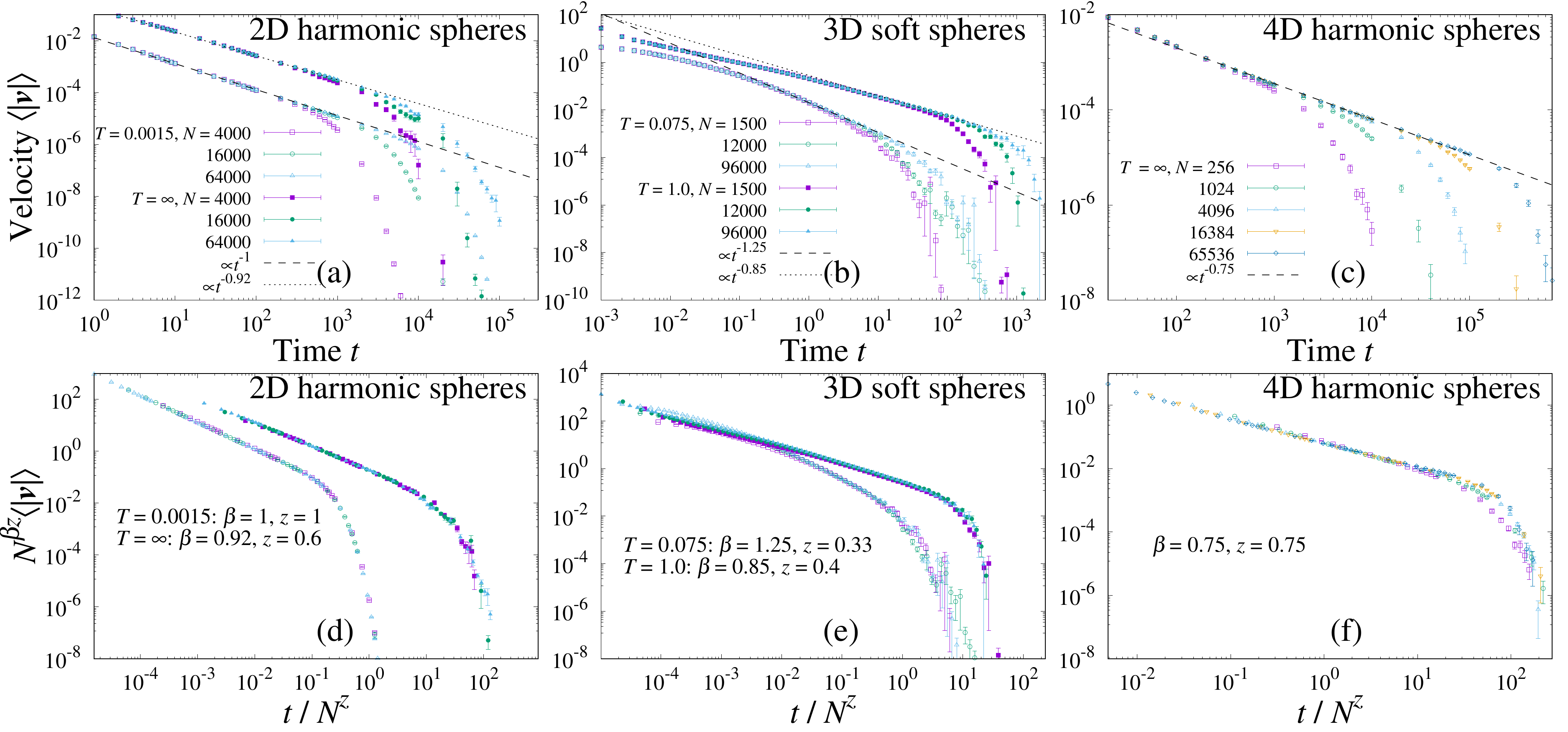}%
\caption{
(a-c) The velocity $\langle |\vec v(t)| \rangle$ as a function of time $t$ 
of the two-dimensional harmonic-sphere, three-dimensional soft-sphere, 
and four-dimensional harmonic-sphere models for several system sizes.
(d-f) The scaling plot using \eq{eq:scaling} for the velocity decays. 
}
\label{fig:v_Ndep}
\end{figure}

In finite size systems, the power-law decay of the velocity is always cut off after a finite time. To conclude whether or not the decay is algebraic in the thermodynamic limit, it is essential to study its system-size dependence. We show $\bigang{|\vec v(t)|}$ of the two-dimensional harmonic spheres, three-dimensional soft spheres, and four-dimensional harmonic spheres for several system sizes in \Fig{fig:v_Ndep} (a-c). In finite dimensions, as we discussed in Sec.II~C, the velocity shows a power-law decay in the short time regime. When the system size increases, the cutoff time becomes longer, meaning that, in the thermodynamic limit $N\to\infty$, the power-law decay continues towards the long-time limit. 
We assume that the velocity decay follows a scaling form
\begin{equation}
\bigang{|\vec v(t)|} = N^{-\beta z}F(t/N^z),
\label{eq:scaling}
\end{equation}
where $F(x)$ is the scaling function, and $z$ is another exponent characterising the cutoff time scale.
Figures~\ref{fig:v_Ndep} (d-f) show the scaling plot using \eq{eq:scaling}.
We find roughly $z \simeq 0.6$ and $0.4$ for velocity decay of the two-dimensional and three dimensional models at high temperature, respectively. In four- and eight-dimensional harmonic spheres in the high-temperature limit and the MK model at high temperature, we observe $z \simeq 0.75$.
At very low temperature, where the harmonic exponent is observed, $z \simeq 1$ and $z \simeq 0.33$ in two and three dimensions, respectively.

\section{Definition of the defect field}

\label{app:defect}

\begin{figure}[b]
\includegraphics[width=.6\linewidth]{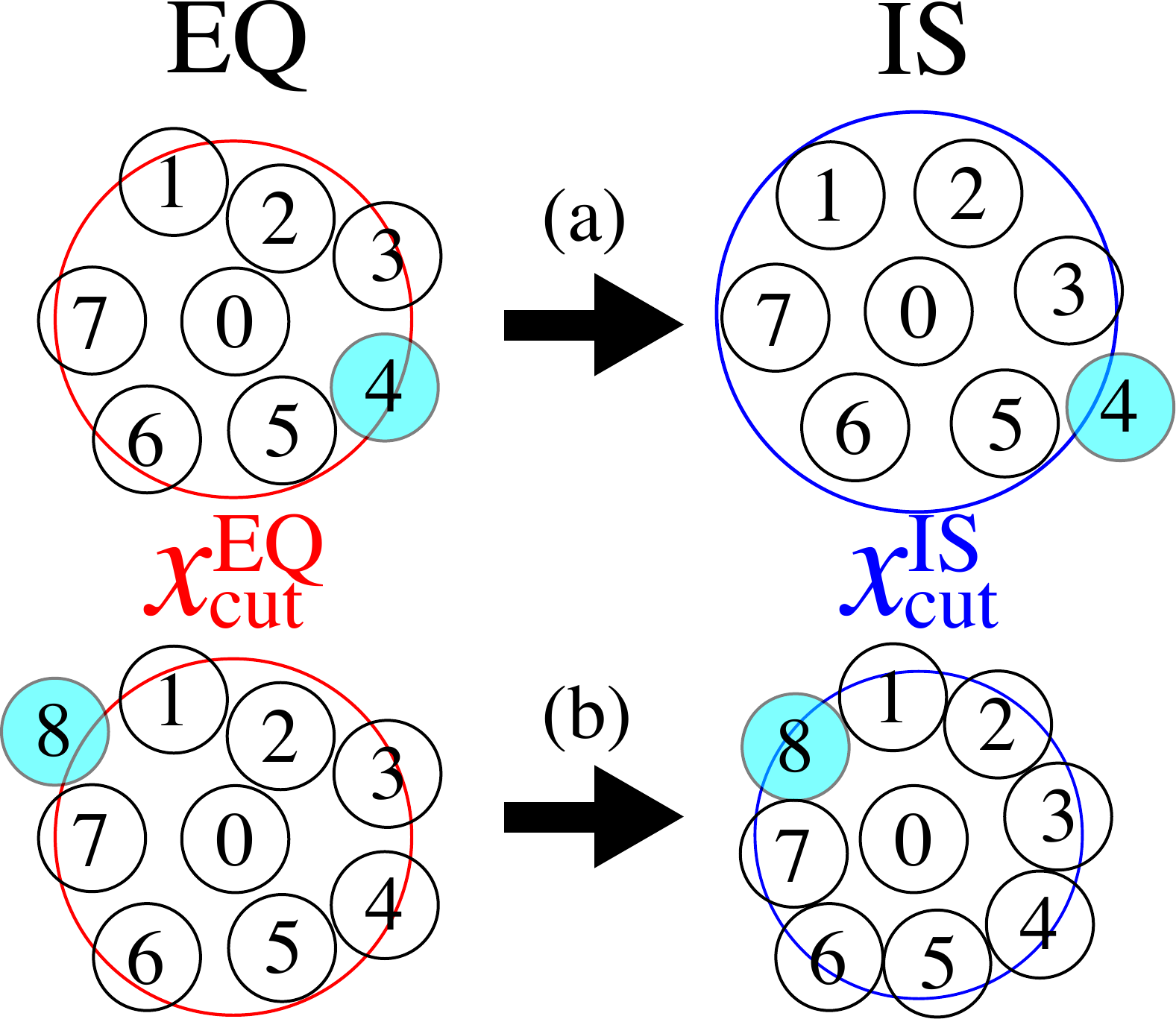}
\caption{Bond-breaking (a) and bond-insertion (b) processes. Schematic pictures show a particle (particle 0) loses a bond (with particle 4) during the minimisation (a), or gains a new bond (with particle 8) (b). Neigbors are defined as a pair of particles closer than a cutoff $x_{\rm cut}^{\rm EQ}$ (which depends on $T$) and $x_{\rm cut}^{\rm IS}$ (which depends on the dynamic process under study).
}
\label{fig:bond}
\end{figure}

\begin{figure}[tbp]
\includegraphics[width=\linewidth]{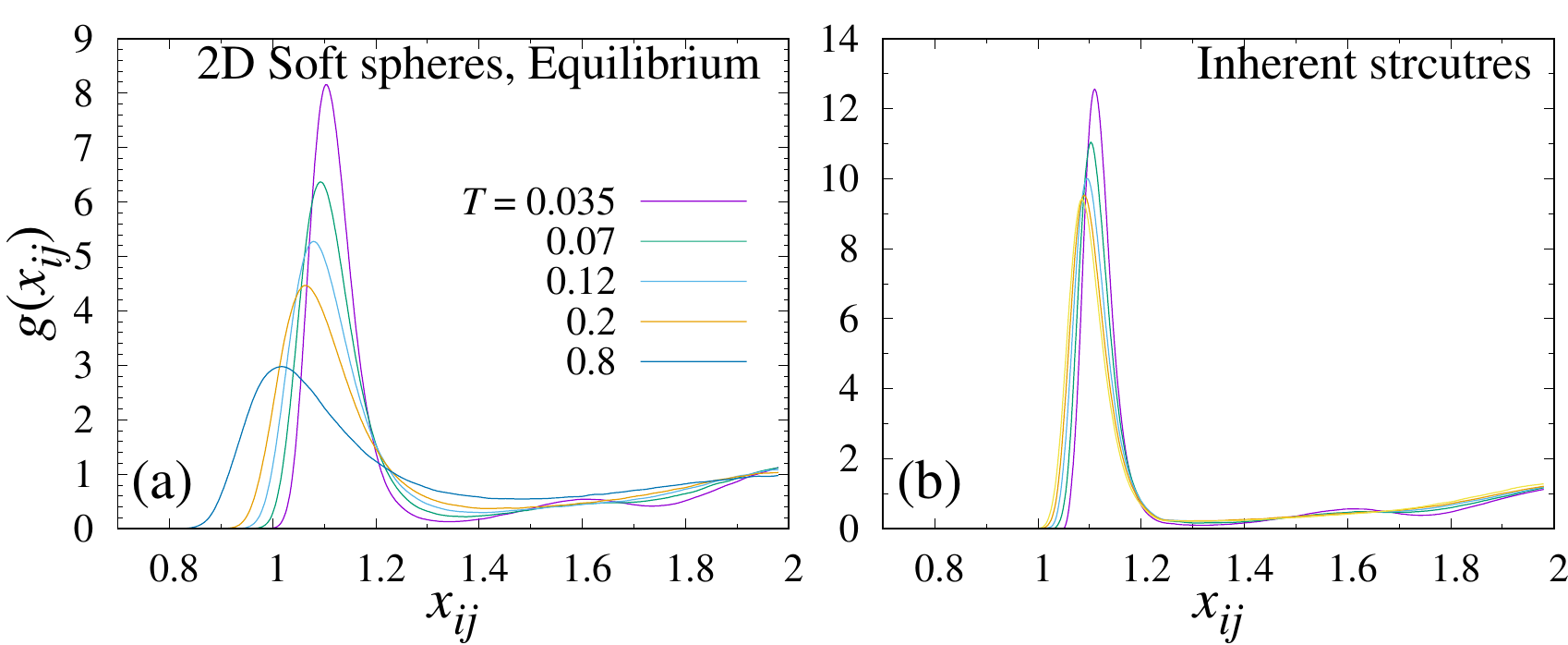}
\caption{
Radial distribution $g(x_{ij})$ as functions of rescaled distances
$x_{ij}=r_{ij}/d_{ij}$ for the equilibrium (a) and inherent structure 
(b) states of the two-dimensional polydispserse soft sphere model. 
The curves are smoothed to determine the first minimum precisely.
The system size $N = 4000$.
}
\label{fig:gofx}
\end{figure}

We explain how to identify defects and measure their concentration. 
We define the defect order parameter $\phi$ based on the rearrangement of neighboring particles between the initial equilibrium and the final inherent structure configurations. Let us first imagine a perfect crystal quenched from a small, finite temperature to zero temperature. The two configurations before and after the quench should be essentially the same except for small thermal fluctuations. In this ideal situation, a given particle neither loses nor gains any neighbour during the quench dynamics. In that case, the defect order parameter should produce zero for all particles. 

Instead, for a disordered liquid at finite temperature, the steepest descent dynamics involves collective rearrangements and a reshuffling of neighboring particles.  When a given particle loses some of the neighboring particles, this process is referred to as a ``bond-breaking'', as studied extensively~\cite{yamamoto1998dynamics,shiba2012relationship}.
We then assign a positive value to the defect order parameter for a particle involved in a bond-breaking process. We also expect that particles can gain new neighbours during the dynamics. Thus, a defect order parameter should also take into account this process which we call a ``bond-insertion''. These two processes are sketched in Fig.~\ref{fig:bond}.

To define neighboring particles in practice, we monitor the radial distribution functions in Fig.~\ref{fig:gofx}. We compute a radial distribution function as a function of a normalized distance, $x_{ij}=r_{ij}/d_{ij}$, which is a suitable representation for continuously polydisperse systems~\cite{Berthier2016}.
In this $2D$ polydisperse model, $g(x)$ for the equilibrium state changes quite a lot because the swap Monte Carlo allows us to sample an extremely wide range of temperatures. Consequently, the first local minimum of $g(x)$, $x_\text{min}^\text{EQ}$, varies $x_\text{min}^\text{EQ} \approx 1.34-1.48$ with varying temperature, $T=0.035-0.800$.
On the other hand, $g(x)$ for the inherent structure barely changes except for the amplitude of the first peak.  In particular, the first local minimum is  located in a nearly flat region ($x \approx 1.23-1.50$).

Based on the observation of $g(x)$, we consider the definition of neighbors for the bond-breaking process. For each particle, we define the neighboring particles in the initial equilibrium configurations as particles located within a cutoff, $x_\text{cut}^\text{EQ}$. We set $x_\text{cut}^\text{EQ}=x_\text{min}^\text{EQ}$. Similarly, we define neighboring particles in the inherent structure by introducing a cutoff $x_\text{cut}^\text{IS}$. Ideally, one would like to use again $x_\text{cut}^\text{IS}=x_\text{min}^\text{EQ}$, but then one finds that some neighbors have a displacement which is just above that threshold, and using the same $x_\text{min}^\text{EQ}$ leads to false neighbor changes. Therefore, one needs to use $x_\text{cut}^\text{IS}>x_\text{cut}^\text{EQ}$ to remove the false positives (see Fig.~\ref{fig:bond}(a)), as already discussed in several previous studies~\cite{yamamoto1998dynamics,shiba2012relationship}. Thus, we set $x_\text{cut}^\text{IS}=1.5$.

Next, we consider the bond-insertion process. 
This can be viewed as a bond-breaking process for the time-reversal of the minimisation dynamics. Applying the same reasoning as above, we impose the condition $x_\text{cut}^\text{IS}<x_\text{cut}^\text{EQ}$ to remove the false positives. Thus we set $x_\text{cut}^\text{IS}=1.23$.

Once neighbors are defined, we determine the list of neighbors for each particle in both equilibrium ($L_\text{eq}$) and inherent structure ($L_\text{IS}$) configurations: $L_\text{eq} = \{ l_\text{eq} \}$ and
$L_\text{IS} = \{ l_\text{IS} \}$, where $l_\text{eq}$ and $l_\text{IS}$ denote the identities of the neighbours. For each particle, say $i$, we define the number of bond breakings, $B_i$ by counting particles with $l_\text{eq} \notin L_\text{IS}$.
We also count the number of bond insertions, $I_i$, which is the number of particles with $l_\text{IS} \notin L_\text{EQ}$.
We note that our choice of cutoff produces nearly identical mean values for  $\overline{B}=\frac{1}{N}\sum_i B_i$ and $\overline{I}=\frac{1}{N}\sum_i I_i$ at all temperatures, meaning that the bond-breaking and bond-insertion processes take place with the same frequency, as expected.

Finally we define the defect order parameter $\phi_i$ as follows: $\phi_i=1$ if $(B_i+I_i)>0$ and $\phi_i=0$ if $(B_i+I_i)=0$. It is useful to make the variable $\phi_i$ binary, as it allows us to determine the absolute concentration of defects, defined as $\phi = \frac{1}{N} \sum_i \phi_i$. In the main text, we report the behaviour of the average defect density, $\langle \phi \rangle$, and of the corresponding collective susceptibility: $\chi = N (\langle \phi^2 \rangle - \langle \phi \rangle^2 ) $.   

\section{Displacement field}

\label{app:disp_field}

\begin{figure}[tbp]
    \centering
    \includegraphics[width=\linewidth]{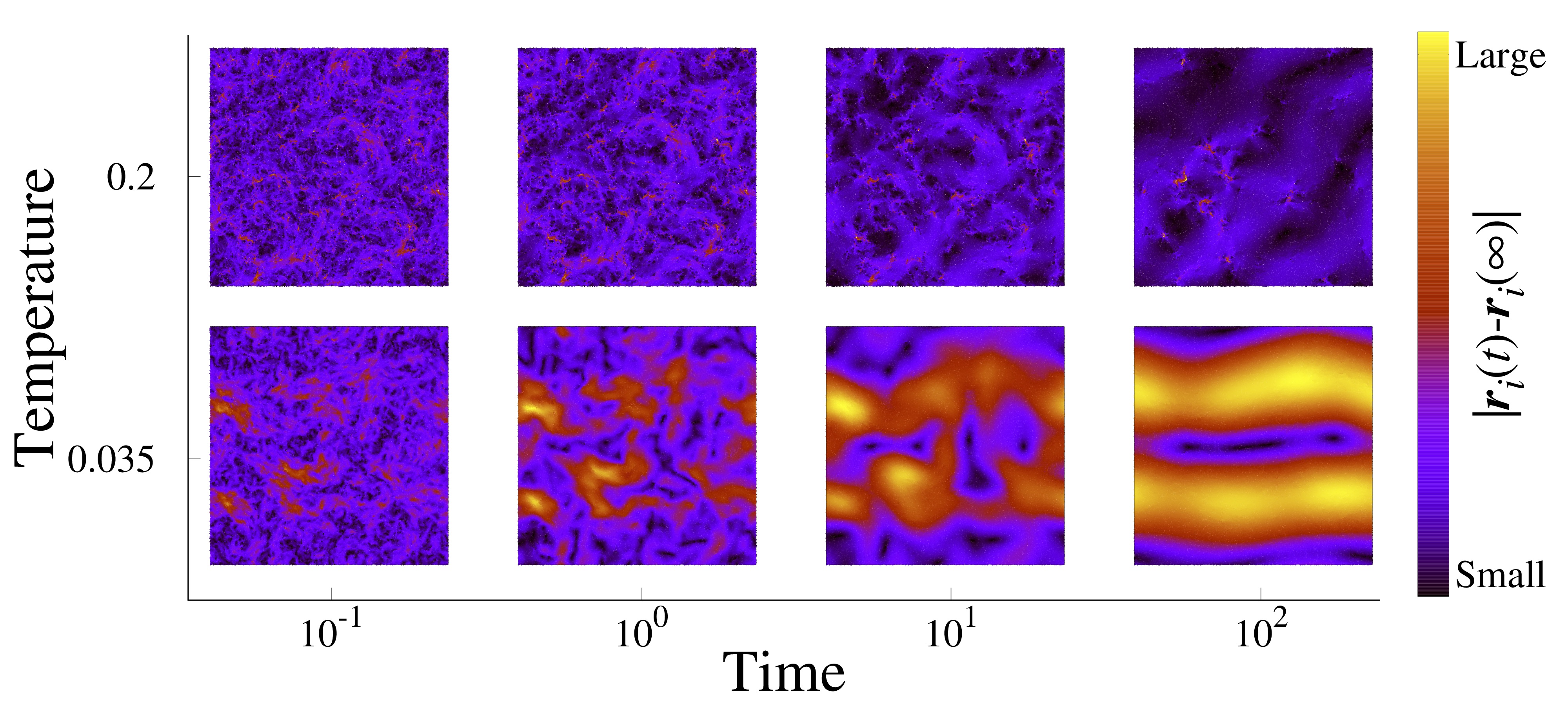}
    \caption{Displacement field $|\vec r_i(t) - \vec r_i(\infty)|$
    for the $2D$ soft sphere model at $T = 0.2$ and $0.035$.
    }
    \label{fig:2dSSdisp_field}
\end{figure}

In \Fig{fig:2dSSdisp_field}, we show the displacement field $|\vec r_i(t) - \vec r_i(t \to \infty)|$ at temperatures $T = 0.2$ and $0.035$ and times $t = 10^{-1}$, $10^0$, $10^1$, and $10^2$. These two temperatures belong to the high- and low-temperature regimes, respectively (see Fig.~2(e)). At high temperature $T = 0.2$, the displacement field reveals many defects at short times. At longer timescales, the number of localised defects decreases, but some of them survive for a very long time. When a defect disappears, particles around the defect rearrange and the velocity $\bigang{|\vec v(t)|}$ has a large sudden increase. The time-evolution of the displacement field at high temperature is consistent with the coarsening picture studied in the $T\to \infty$ limit for harmonic spheres in Ref.~\cite{Chacko2019}.

At lower temperature $T=0.0035$, on the other hand, the displacement field is much smoother even at short times, and we hardly see localised defects. This smooth displacement field is consistent with a harmonic description, as shown in Sec.II~C.

\subsection{Inherent structures from different algorithms}

\label{app:SDCGFIRE}

\begin{figure}[t]
\includegraphics[width=.5\linewidth]{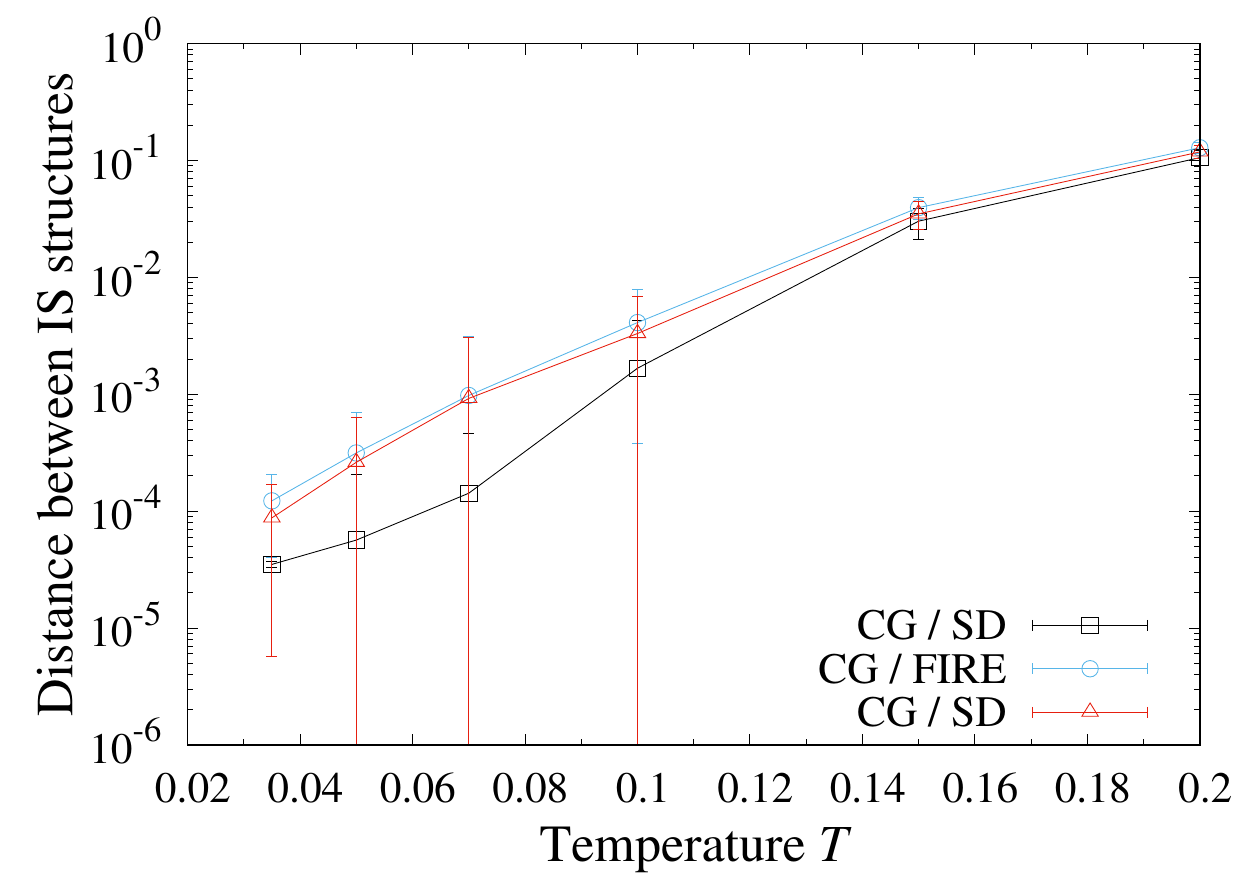}
\caption{
The distance between the inherent-structure configurations of the two-dimensional 
soft-sphere model produced by the steepest descent (SD), conjugate gradient (CG), and FIRE algorithms \cite{Bitzek2006}. 
$N=64000$.}
\label{fig:SDCGFIRE}
\end{figure}

We show that localised defects affect the mapping between an equilibrium liquid microstate and a `corresponding' minimum in the energy landscape, i.e. an inherent structure. Starting from the same initial equilibrium state,  we minimize the energy using different algorithms available in LAMMPS \cite{plimpton1995fast}, namely conjugate gradient, FIRE~\cite{Bitzek2006}, and steepest descent. This exercise is done for the $2D$ system of polydisperse soft spheres. In all cases, we use the same tolerance thresholds (energy tolerance of $10^{-16}$ and force tolerance of $10^{-20}$ \footnote{The energy tolerance is defined as $\Delta{E}/E$, where $E$ is the total energy of the system and $\Delta{E}$ is the energy difference between successive iterations during minimization. The force tolerance is defined as the length of the global force vector for all atoms, e.g. a vector of size $3N$ for $N$ atoms. \cite{plimpton1995fast}}) to end the convergence to a local minimum. After obtaining the minima corresponding to the same initial state via the three different algorithms, we compute the distance between the minima obtained for each pair of these algorithms, which is given by
$\left\langle \sqrt{\frac{1}{N} \sum_{i=1}^N|\vec{r}_i^{\rm A}-\vec{r}_i^{\rm B}|^2} \right \rangle$,
where $\vec{r}_i$ is the co-ordinate of the $i$-th particle in the configuration, and A,B represent labels corresponding to CG, SD, or FIRE.
If the pair of minima are identical, this distance should be very close to zero (and is only set by the convergence criterion). 

However, this is not what we observe, as shown in Fig.~\ref{fig:SDCGFIRE}. Instead we find that the average distance is not given by the convergence criterion (it is much larger), and it depends continuously upon the temperature from where the initial state is sampled; it is larger for higher temperatures.

Only at very low temperatures in finite systems do we find that instances where the pair of minima obtained from different algorithms can become similar, within numerical resolution. In those cases, the initial equilibrium state rolls down to the same minimum, independent of the algorithm used and the mapping between an equilibrium liquid state and a local minimum in the energy landscape becomes unique. However, when increasing the system size and/or the temperature, these minima are always distinct and we suspect that this is always the case in the thermodynamic limit at any finite temperature. 
The difference between pairs of minima obtained from a given initial configuration is visualized in the main text.

\bibliography{refs}